\documentclass[aps,twocolumn]{revtex4}

\begin{document}
\draft
\title{Controlling Polar Molecules in Optical Lattices.}

\author{S. Kotochigova$^{a,b}$ and E. Tiesinga$^{b}$}

\address{$^{a}$ Department of Physics, Temple University, Philadelphia, PA 19122, \\
$^{b}$ National Institute of Standards and Technology, 100 Bureau Drive,
stop 8423, Gaithersburg, Maryland 20899.}

\begin{abstract}
We investigate theoretically the interaction of polar molecules with
optical lattices and microwave fields. We demonstrate the existence
of frequency windows in the optical domain where the complex internal
structure of the molecule does not influence the trapping potential of the
lattice.  In such frequency windows the Franck-Condon factors are so small
that near-resonant interaction of vibrational levels of the molecule with
the lattice fields have a negligible contribution to the polarizability
and light-induced decoherences are kept to a minimum.  In addition, we
show that microwave fields can induce a tunable dipole-dipole interaction
between ground-state rotationally symmetric ($J$=0) molecules. 
A combination of a carefully chosen lattice frequency and microwave-controlled
interaction between molecules will enable trapping of polar molecules
in a lattice and possibly realize molecular quantum logic gates.
Our results are based on {\it ab~initio} relativistic electronic
structure calculations of the polar KRb and RbCs molecules combined with
calculations of their rovibrational motion.

\end{abstract}

\maketitle
Quantum gases of ultracold atoms and molecules, confined in
periodic optical lattices, open exciting prospects for ultimate control 
of their internal and external degrees of freedom. This control can be achieved
by changing intensity, wave length, and geometry of the laser fields that
create the lattice.
Spectacular progress has been made in loading and manipulating atomic 
species in periodic optical potentials \cite{Meacher,Greiner,Grynberg,Monroe,Bloch}.  
Theoretical developments have played an important role in this progress 
leading to a better fundamental understanding of observed phenomena and 
probing the possibility to use optical lattices in quantum computation 
with ultracold atoms \cite{Jessen,Jaksch}. 

Currently, the goal of many researchers is to perform similar experiments
on cold molecular species.  Optical lattices might be filled with
ultracold molecules which have been formed by photoassociation from
ultracold colliding atoms \cite{Julienne,Wynar,Rom} or alternatively be loaded
with molecules cooled by deceleration \cite{Meijer} or thermalization
with He \cite{Doyle}. Developments with trapped cold molecules
have started a new field of physics \cite{Review}.

Polar molecules are of particular interest in such experiments.  This interest is
based on the fact that polar molecules have a permanent dipole moment and
therefore can interact via long-range dipole-dipole interactions. Trapped
in an optical lattice the dipole-dipole interactions can make new types
of highly-correlated quantum many-body states possible \cite{Damski}.
Moreover, it has been proposed \cite{DeMille} that polar molecules
trapped in an optical lattice can be good candidates for quantum
bits of a scalable quantum computer.  The electric dipole moment of
ultracold polar molecules oriented by an external electric field may
serve as the quantum bits and entanglement between qubits is due to
the dipole-dipole interaction \cite{Raussendorf} between molecules.
In principle, dipole moments can be induced in atoms and homonuclear
molecules by external electric fields. For experimentally feasible static
fields these induced dipole moments are weak on the order of $10^{-2}$
D \cite{Townes}. The permanent dipole moments in polar molecules
are much larger with typical values of 1 D.

In this Letter we investigate theoretically the effects of optical lattice
and microwave fields on polar molecules.  
The former are used to trap and the latter to align molecules.
We start from ultracold polar molecules in the lowest rovibrational
level of the ground electronic potential, possibly created by Raman
photoassociation from its individual atoms. Moreover, we assume 
that there is one molecule per a lattice site. Here we focus on KRb
and RbCs polar molecules as they are proposed to be good candidates
for ultracold experiments \cite{Stwalley} and qubits of quantum
information devices  \cite{DeMille}. 

Localization of the molecule in a lattice site requires sufficiently
deep lattice potentials, at least tens of kHz,  to prevent tunneling of the molecule from one
site to another. This can be achieved by increasing the laser intensity.
On the other hand, higher laser intensities  increase
the possibility of exciting a molecule leading to unwanted decoherence
processes. In particular, lattice fields of specific frequencies in the optical domain
can transfer population from the lowest rovibrational level of the
ground potential to a rovibrational level of an excited potential,
which then by the spontaneous emission can decay to many rovibrational
levels of the ground potential. As result, we lose control over the
molecule in the lattice. One of our goals here is to determine lattice parameters
that ensure strong trapping forces and simultaneously lead to small
decoherence rates.  

Controllable dipole-dipole interaction between polar molecules in
a lattice lies at the heart of proposals to exploit entanglement
as an essential resource for strongly correlated many-body states and
quantum information processing.  Even though polar molecules have a
permanent electronic dipole moment, for any $J=0$ rotational state it
averages to zero. Other rotational states ($J \geq$ 1) do have a non-zero
permanent dipole moment.  In fact, for the $v=0$ level of the ground electronic
state it is 0.76 D
for KRb \cite{Kotochigova1} and 1.27 D for RbCs \cite{Kotochigova}.
We propose that a dipole moment is
easily induced in a $J=0$ state by applying a microwave field with a  frequency
that is close to the $J = 0$ to $J = 1$ resonance. Moreover, we investigate conditions 
under which the relevant states for a microwave transition have exactly the
same AC Stark shift to minimize the field perturbations.

The relevant property for controlling a molecule with a light field is the 
complex molecular dynamic polarizability $\alpha(h\nu,\vec{\epsilon})$ as
a function of radiation frequency $\nu$ and polarization $\vec{\epsilon}$
($h$ is Planck's constant).
Assuming that the alkali-metal molecule is in a rovibrational state of the 
ground X$^1\Sigma^+$ potential, its dynamic polarizibility in SI units is
given in terms of the dipole coupling to other rovibrational states of the ground
and excited potentials as
\begin{eqnarray}
   \lefteqn{ \alpha(h\nu,\vec{\epsilon}) = } 
      \label{eqpolar}
\\
  &&\frac{1}{\epsilon_0c}
   \sum_{f} \frac{(E_f - ih\gamma_f/2 - E_i)}{(E_f - ih\gamma_f/2 - E_i)^2 - (h\nu)^2}
     \times |\langle f|d \,\hat{R}\cdot \vec{\epsilon}|i\rangle|^2. \nonumber
\end{eqnarray}
where $c$ is the speed of light, $\epsilon_0$ is the electric constant,
$\hat{R}$ is the orientation of the interatomic axis, $i$ and $f$ denote 
the initial $|vJM\rangle$ and intermediate $|v'J'M'\rangle$
rovibrational wavefunctions of the $|X^1\Sigma^+\rangle$ and $|\Omega\rangle$
electronic states, respectively.
Here, $\Omega$ labels either the X$^1\Sigma^+$
state or any excited state and  $\langle f|d|i \rangle$
are $R$-dependent permanent or transition electronic dipole moments.
The quantities $M$ and $M'$ are the projections along a laboratory
fixed coordinate system of $\vec{J}$ and $\vec{J}'$, respectively.
The energy $E_i$ is a rovibrational energy in the X$^1\Sigma^+$
state and $E_f$ is the rovibrational energy of the intermediate
$\Omega$ states.  Finally, the line widths $\gamma_f$
describe the spontaneous and any other decay mechanism that leads
to loss of molecules.  Equation~\ref{eqpolar} includes a sum over
dipole transitions to rovibrational levels within the
X$^1\Sigma^+$ potential as well as to rovibrational levels of excited
potentials.  Contributions from
scattering states or the continuum of the  $\Omega$ states must also be
included. The sum excludes the initial state. This sum, however, can be truncated assuming that transitions may have zero or near zero electronic dipole moments and/or are far
detuned.  Nevertheless, a significant number of excited potentials and
vibrational states have to be included.

\begin{figure}
\vspace*{6.0cm}
\includegraphics{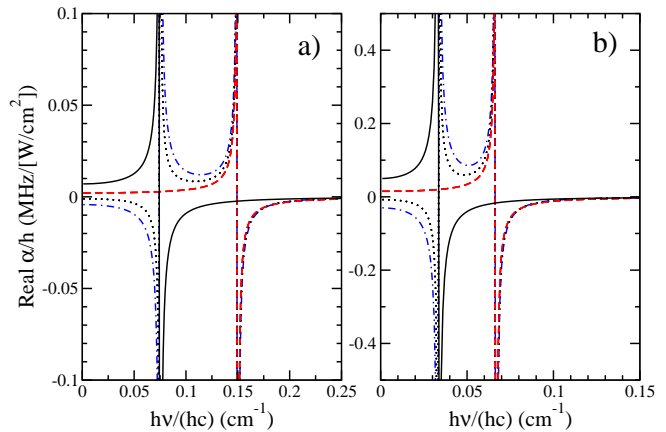}
\caption{Dynamic polarizability for various photon polarizations of KRb
(panel a) and RbCs (panel b) in the $J=0$ or $J=1$, $v=0$ rovibrational
level of their X$^1\Sigma^+_0$ electronic ground state as a function of
the microwave frequency. The solid line in both panels corresponds to the
polarizibility of the $J=0$ level, which is independent of the photon polarization.  
All other curves are for the $J=1$ level.
The dotted line corresponds to the polarizibility of M=$\pm$ 1 magnetic
sublevels illuminated by linear polarized $\sigma_x$ or $\sigma_y$ radiation. 
The dashed
line relates to either M=$\pm$ 1 sublevels illuminated by $\sigma_z$
light or the $M=0$ level with $\sigma_x$ or $\sigma_y$ light.  Finally,
the dashed-dotted line is obtained for $M=0$ levels with $\sigma_z$ light.
Note that the scale on both axes differs for the two panels.  }
\label{J01_micro}
\end{figure}

The calculation of the molecular polarizability requires knowledge
of molecular potential surfaces, permanent and transition dipole moments.
We use data from our previous electronic structure
calculations of KRb and RbCs polar molecules \cite{Kotochigova,
Kotochigova1,Kotochigova2}. Here we combined them with a nuclear 
dynamics calculations,  based on a discrete variable representation 
\cite{Tiesinga},  to obtain  the rovibrational energies of and Franck-Condon
factors between the ground and excited potentials of KRb and RbCs.  

Our results for the molecular polarizability of KRb  and RbCs at 
very-long wavelengths corresponding to microwave frequencies are
shown in Fig.~\ref{J01_micro}. 
Unlike the $J=0$ rotational level, the polarizability
of J$\geq$ 1 rotational levels depends on the polarization of light
and  the projection of $\vec{J}$.  For microwave frequencies electric
dipole transitions within the ground X$^1\Sigma^+$ potential dominate
in the sum of Eq.~\ref{eqpolar}.  This X$^1\Sigma^+$ contribution does
not exist for homonuclear molecules.  The resonances in the graph are
due to the rotational transitions from $J = 0$ to $J = 1$ and from $J =
1$ to $J = 0$ and $J = 2$ within the same $v=0$ vibrational state.
The $J = 1$ to $J = 2$ transition occurs at a larger photon frequency.
For the near-resonance frequencies the polarizabilities
in Fig.~\ref{J01_micro} are approaching 1 MHz/(W/cm$^2$), which 
is much larger than atomic polarizabilities of K, Rb, and Cs in the same frequency region. 
For example, the dynamic polarizability of a Rb atom \cite{Safronova}
is 4 to 5 orders of magnitude smaller than the dynamic polarizability
of the RbCs molecule at the same frequency.  

Figure~\ref{J01_micro} also shows that the $J=0$ and $J=1$
polarizabilities generally differ and even have
opposite signs. However, although the vertical scale of the figure is
too large to see this, for the X$^1\Sigma^+$ $v=0$ level they are the
same at a frequency of $h\nu/(hc)\approx$ 0.3 cm$^{-1}$ for KRb and 0.1
cm$^{-1}$ for RbCs.  These might be so-called magic frequencies at which
a microwave field creates the same AC Stark shift for
a $J=0$ and $J=1$ state. Our estimate shows that at the magic frequencies $Re~\alpha/h$
equals to $-$0.5 kHz/(W/cm$^2$) for KRb and $-$3 kHz/(W/cm$^2$) for RbCs.

For microwave fields nearly resonant with the $J =0$ to $J = 1$ transition  the $J = 0$ 
molecular state is ``dressed'' and acquires a dipole moment.
In fact, from perturbation theory the strength of the
induced dipole moment of the ``dressed'' $J = 0$ state is
\begin{eqnarray} 
\left|\vec d_{\rm ind} \right|
&\approx&
    \left|
  2 \frac{ \langle\langle v,0 | -\vec d\cdot\vec{\cal E} |v,1\rangle\rangle  }
         {h\nu-(E_{X,v,J=1}-E_{X,v,J=0})}
        \langle\langle v,0 | \vec d | v,1 \rangle\rangle \right|
  \nonumber \\
&\sim&
    \sqrt{4\pi\epsilon_0\frac{c}{2\pi}} \alpha(h\nu,\vec\epsilon)\sqrt{I}
\end{eqnarray} 
proportional to its polarizibility.
Here $|v,J\rangle\rangle=|X^1\Sigma^+\rangle|v,J M\rangle$, the
energy $-\vec{d}\cdot \vec{{\cal E}}$ is the molecule-light coupling,
$\vec {\cal E}$ the applied electric field, and $I$ the corresponding intensity.  The projection of
the $J$ state is determined by the polarization of the microwave field.
Our estimates for the $v=0, J=0$ rovibrational level of RbCs indicate that
in the limit of zero frequency, where $Re~\alpha/h$ = 0.05 MHz/(W/cm$^2$),
and $I$ = 100 W/cm$^2$ the induced
dipole moment is on the order of 0.05 times the permanent dipole moment,
$\langle\langle v,1 | \vec d | v,1 \rangle\rangle$, of the $v=0, J=1$ level. 
At the magic microwave frequency a field intensity
larger than 100 W/cm$^2$ is needed to achieve appreciable induced dipole moments.

For a microwave field resonant with the $J =
0$ to $J = 1$ transition  the  ``dressed'' $J = 0$ molecular state 
has an induced dipole moment of one half of the
permanent dipole moment of the $v=0, J=1$ state. Assuming that
$J=0$ polar molecules in neighboring lattice sites interact via tunable
dipole-dipole forces V$_{dd}\sim$ $d_{ind}^2$/$R_L^3$, where $R_L$ is
one half of a lattice wave length, we can determine the characteristic
interaction time $\delta t$ = $h/V_{dd}$. If the lattice is
made from  laser light at optical frequencies, say at 690 nm for KRb or 810 nm for RbCs
(we will justify this choice of frequencies later on), the interaction
time for these molecules is as short as 2 ms. This time is
short enough to establish entanglement of molecules during ultracold
experiments with typical time scales of a few hundred milliseconds.

Light-induced decoherence of rovibrational levels of a polar molecule in
the microwave region has two sources.  The first source is spontaneous
emission by electric dipole transitions to lower lying rovibrational
levels.  More important, however, is the decoherence due to black-body
radiation from the room-temperature environment in the experiments.
These two processes combined lead to lifetimes that are long ($\geq$
100~s as was shown in Refs.~\cite{Kotochigova2,Kotochigova}) compared
to  currently realistic experimental time scales. 

\begin{figure}
\vspace*{6.3cm}
\includegraphics{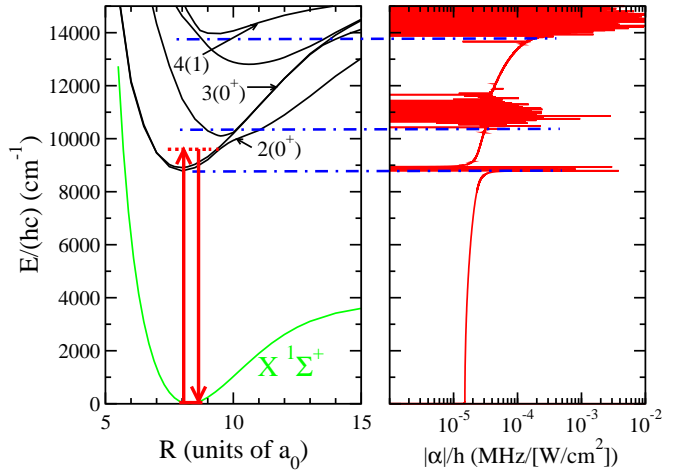}
\caption{The ground and lowest excited potentials of RbCs as a
function of internuclear separation (left panel) and the absolute
value of the dynamic polarizability of the $v=0$ and $J=0$ level of the
X$^1\Sigma^+$ electronic ground state as a function of laser frequency
(right panel). The zero of energy of the left panel is the energy
of the $v=0,J=0$ X$^1\Sigma^+$ level. The vertical axis of the two
panels is the same. The two arrows illustrate a contribution to the
dynamic polarizibility. Three excited potentials with relativistic
symmetry $\Omega^\pm=0^+$ have been labeled. The horizontal dashed
lines correlate features in the polarizability with features in the
potentials. (The polarizibility is only evaluated at frequencies spaced
by $h\Delta\nu/(hc)=1$ cm$^{-1}$.)}
\label{pot-pol} 
\end{figure}

In the infrared and visible part of the spectrum molecules can be trapped
in individual sites of the optical lattice.
The depth of the lattice potential
is determined by the real part of polarizability via $V_0 = -Re~
\alpha(h\nu,\vec{\epsilon})\times I$, where $I$ is the intensity of
a field at frequency $\nu$ and polarization $\vec{\epsilon}$,
while the laser-induced decoherence is proportional to the imaginary
part of the polarizability.
The left panel of Fig.~\ref{pot-pol} shows the
potential energy curves of RbCs that are most relevant for and included
in our calculation of the polarizibility of rovibrational levels of the
X$^1\Sigma^+$ state at these laser frequencies.  The right panel shows the
absolute value of the polarizibility of the $v=0,J=0$ rovibrational level
of the X$^1\Sigma^+$ state of RbCs.  For infrared and optical domains
the dominant contribution to $\alpha$ is from rovibrational levels of
the excited electronic potentials rather than from the ground state
potential.  For KRb the potential energy curves and the polarizibility
have a similar structure.

Figure~\ref{pot-pol} shows that the polarizabilitiy has frequency windows 
where it is a slowly varying or ``static'' function  interspersed with
regions of multiple closely-spaced resonant-like features, where
$\alpha$ can be orders of magnitude larger.  The resonant-like features
are due to bound states of excited potentials and are closely
related to inner- and outer-turning points of the excited
potentials in the left panel of Fig.~\ref{pot-pol}.  The vibrational
wavefunction of the $v=0,J=0$ X$^1\Sigma^+$ level is highly localized
around the minimum of the X$^1\Sigma^+$ potential and, hence, a large
vibrationally averaged transition dipole moment occurs when both the
electronic dipole moment is large and excited vibrational levels have good
overlap with the $v=0,J=0$ X$^1\Sigma^+$ level. Dash dotted horizontal
lines in Fig.~\ref{pot-pol} make this connection between the two panels.

In the slowly varying regions of
$\alpha$ above 9000 cm$^{-1}$ there are hidden resonances, that have
negligible contribution to the polarizability due to small Franck-Condon 
factors with the ground state. For the figure,
the polarizibility is evaluated every $h\Delta\nu/(hc) = 1$ cm$^{-1}$
or $\Delta\nu$ = 30 GHz, which is large compared to the vibrational level linewidths
$\gamma\sim$ 6 MHz. This implies that if the vibrationally
averaged transition dipole moment is large compared to or on the order
of $h\Delta\nu$ a resonance appears in Fig.~\ref{pot-pol} while if it is small no resonance
is visible. 

\begin{figure}
\vspace*{7cm}
\includegraphics{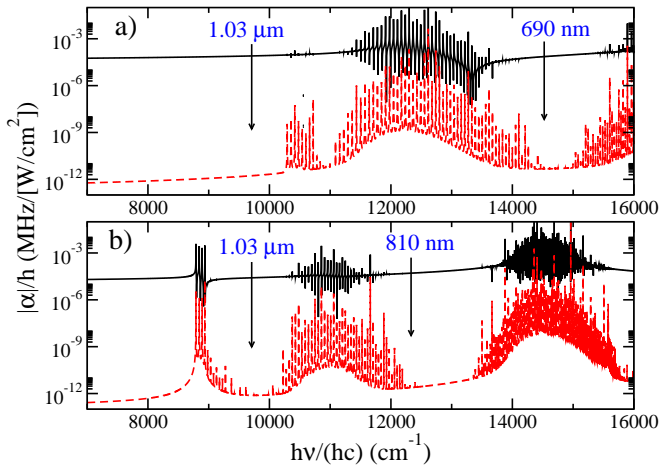}
\caption{Real (solid line) and imaginary part (dashed line) of the dynamic
polarizability of the $v=0,J=0$ level of the X$^1\Sigma^+$ state
of KRb (panel a) and RbCs (panel b) as a function laser frequency. Practical
laser frequencies are indicated.}
\label{real-imag}
\end{figure}

Figure~\ref{real-imag} shows both real and imaginary part of the
polarizability of the $v=0, J=0$ state of the X$^1\Sigma^+$ potential
of KRb (upper panel) and RbCs (lower panel) in the optical domain.
The imaginary part is proportional to the line width of excited states
and is smaller than the real part of polarizability for both molecules.
In fact, the ratio between real and imaginary part of the polarizability
is about 10$^7$ away from the resonances and significantly smaller ($\sim
10^3$) near them.  Optical potentials seen by KRb or RbCs molecules in these
regions can be very deep ($V_0/h$ $\approx$ 1 MHz) for laser
intensities of the order of 10$^4$ W/cm$^2$. At such intensities tunneling
of molecules from one lattice site to another is negligible. Moreover,
the decoherence time is significantly larger than 1 s.  

Using our results we propose two frequency intervals in which resonant
excitation is unlikely and are most easy to work with experimentally. 
For the lattices in an optical domain 
we suggest lasers with wavelengths between
680$\pm$35 nm for a KRb and 790$\pm$40 nm for a RbCs experiment. In
addition, telecommunication wavelengths between 1.03$\pm$0.05 $\mu$m
seem practical for both molecules.

We conclude by saying that despite the fact that molecules have an
internal structure that is more complex than atoms, this complexity can
be used to good advantage or neutralized. Microwave fields nearly resonant with rotational
transitions within the molecule can lead to tunable interactions between 
neighboring molecules.
For optical lattice potentials created by standing light
waves a careful selection of laser frequency in regions, where
the Franck-Condon factors to excited vibrational levels are small, creates conditions
for which the ratio of coherent to decoherent effects is large and nearly independent
on the internal molecular structure.
Moreover, heavy polar alkali metal molecules can be strongly confined in an 
optical lattice with relatively modest intensities.

The work was supported in part by the Army Research Office.  We acknowledge
useful discussions with Dr. J. V. Porto and Profs. D. DeMille and
S. L. Rolston.

\end{document}